\documentstyle[epsf,12pt]{article}
\begin{document}

\begin{center}
{\bf TOWARDS AUTOMATIC ANALYTIC EVALUATION
OF DIAGRAMS WITH MASSES
\footnote{~ Basic points of this paper have been briefly presented
at The International Workshop AIHENP '96 \cite{our}}.}

Leo.~V.~Avdeev
\footnote{~E-mail: avdeevL@thsun1.jinr.dubna.su}
J.~Fleischer
\footnote{
~Fakult\"at f\"ur Physik, Universit\"at Bielefeld,
D-33615 Bielefeld 1, Germany.
\\
E-mail: fleischer@physik.uni-bielefeld.de},~
M.~Yu.~Kalmykov
\footnote{~E-mail: kalmykov@thsun1.jinr.dubna.su}
and M.~N.~Tentyukov
\footnote{~E-mail: tentukov@physik.uni-bielefeld.de}

\vspace{15mm}

{\it Bogoliubov Laboratory of Theoretical Physics,
Joint Institute for Nuclear Research,
Dubna, Moscow Region 141980,
Russian Federation}

\end{center}

\begin{abstract}
A method to calculate two-loop self-energy diagrams of the Standard
Model is demonstrated. A direct physical application is the
calculation of the two-loop electroweak contribution to the anomalous
magnetic moment of the muon ${\frac{1}{2}(g-2)}_{\mu}$. Presently, we
confine ourselves to a ``toy'' model with only $\mu$, $\gamma$ and a
heavy neutral scalar particle (Higgs).  The algorithm is implemented
as a FORM-based program package. For generating and automatically
evaluating any number of two-loop self-energy diagrams, a special
C-program has been written. This program creates the initial
FORM-expression for every diagram generated by QGRAF, executes the
corresponding subroutines and sums up the final results.
\end{abstract}

\vspace*{10pt}

{\it PACS number(s)}: 12.15.Lk, 13.10.+q, 13.40.Em

anomalous magnetic moment

\pagebreak
\section{Introduction}
\vspace*{-0.5pt}

   Recent high precision experiments to verify the Standard Model of
electroweak interactions require, on the side of the theory,
high-precision calculations resulting in the evaluation of higher
loop diagrams.  For specific processes thousands of multiloop Feynman
diagrams do contribute, and it turns out impossible to perform these
calculations by hand. That makes the request for automatization a
high-priority task. In this direction, several program packages are
elaborated \cite{GRACE,FeynmArts,CompHep,Vermaseren}.  It appears
absolutely necessary that various groups produce their own solutions
of handling this problem: various ways will be of different
efficiency, have different domains of applicability, and last but not
least, should eventually allow for completely independent checks of
the final results. This point of view motivated us to seek our own
way of automatic evaluation of Feynman diagrams. We have in mind only
higher loop calculations (no multipoint functions).

There exist several kinds of methods for evaluating multiloop Feynman
diagrams with masses.  The most direct and universal method is based
on the Feynman parameter representation and subsequent numerical
Monte-Carlo integration \cite{Fujimoto}. The universality is its most
essential advancement.  Great progress has been achieved in
implementing this method into combined automatic systems like GRACE
\cite{GRACE}. But the convergence of the Monte-Carlo integration is
rather slow, and there is no way to estimate the actual numerical
error. When the subintegral expression has kinematical singularities,
the error may significantly exceed the estimate \cite{MC-error}.

In a semi-analytic method some integrations are done exactly in terms
of special functions, and the dimension of numerical integration
becomes lower \cite{Kreimer}, \cite{self-energy}.  The advantage of
this method is the possibility to deal with tensor structures.
However, an essential drawback is the difficulty of consistently
performing renormalizations, because one has to stay in the integer
dimension.

The method \cite{Pade} of Taylor expansion of a diagram in external
momenta, analytic continuation and Pad{\'e}-like approximations
allows one to recover the behaviour of the function in the whole
complex plane of momentum variables. However, it would require the
knowledge of rather many expansion coefficients in order to get a
sufficiently precise estimate at the threshold.

Therefore, we are going to use the asymptotic expansions of Feynman
diagrams in small/large momenta and masses, where the orders of
expansion in a small ratio of parameters are completely collected
even at the threshold. We are interested in the low-energy physics of
the Standard Model. In this situation there exists a natural small
parameter: the ratio of the characteristic scale of the process to
the scale of the weak interaction, defined by $M_Z$.  This provides a
good convergence for asymptotic expansions of physical observables
and, in a number of cases, makes already the leading order sufficient
for the existing precision of the experiments.  On the other hand,
one can use the dimensional renormalization, and then there are no
problems with the $R$ operation.

We demonstrate here the functioning of a C program (TLAMM) for the
evaluation of the two-loop anomalous magnetic moment (AMM) of the
muon ${\frac{1}{2}(g-2)}_{\mu}$.  This piloting C program must read
the diagrams generated by QGRAF \cite{QGRAF} for a given physical
process, generate the FORM \cite{FORM} source code, start the FORM
interpreter, read and sum up the results for the class of diagrams
under consideration.  Here, for the purpose of demonstration, we
apply TLAMM to a closed subclass of diagrams of the Standard Model
which we refer to as a ``toy'' model.

Recent papers have reduced the theoretical uncertainty of the muon
AMM by partially calculating the two-loop electroweak contributions
\cite{Kuraev}, \cite{Czarnecki}.  In some cases the following
approximations were used:  terms suppressed by $(1-4 \sin^2 \Theta_W
)$ were omitted; the fermion masses of the first two families were
set to zero; diagrams with two or more scalar couplings to the muon,
suppressed by the ratio $\frac{m^2_{\mu}}{M^2_Z}$, were discarded;
the Kobayashi-Maskawa matrix was assumed to be unity; the mass of the
Higgs particle was assumed large as compared to $M_{Z}$.

All of these approximations, except possibly the last one, are well
justified and give rise to small corrections only. We consider it of
great interest to study also the case $M_H \sim M_{Z}$.  To perform
this calculation is our main physical motivation. Apart from that,
for technical reasons, it may be interesting to study the functioning
of TLAMM by calculating all $1832$ two-loop diagrams without any
approximation.

The calculation of the AMM of the muon reduces, after differentiation
and contractions with projection operators, to diagrams of
the propagator type with external momentum on the muon mass shell
(for details see \cite{projector}).

The applicability of the asymptotic expansion \cite{asymptotic},
\cite{Smirnov} in the limit of large masses has to be investigated for
diagrams evaluated on the muon mass shell ({\em i.e.},
$p^2=-m_{\mu}^2$). Some diagrams already had a threshold at the muon
mass shell before the expansion.  In other diagrams, this threshold
appears in some terms of the expansion.  In dimensional
regularization, threshold singularities (like any other infrared
singularities if they are strong enough) manifest themselves as poles
in $\varepsilon$ (in $4-2\varepsilon$ dimensions). They ought to
cancel for the total AMM. We check this for a closed subset of
diagrams in the toy model.

\section{Large-mass expansion}

The asymptotic expansion in the limit of large masses is defined
\cite{Smirnov} as

\begin{equation}
F_G(q, M ,m, \varepsilon) \stackrel{M \to \infty}{\sim }
\sum_{\gamma} F_{G/\gamma}(q,m,\varepsilon) \circ
T_{q^{\gamma}, m^{\gamma}}
F_{\gamma}(q^{\gamma}, M ,m^{\gamma}, \varepsilon)
\end{equation}

\noindent
where $G$ is the original graph, $\gamma$'s are subgraphs involved in
the asymptotic expansion, $G/\gamma$ denotes shrinking $\gamma$ to a
point; $F_{\gamma}$ is the Feynman integral corresponding to
$\gamma$; $ T_{q_{\gamma}, m_{\gamma}} $ is the Taylor operator
expanding the integrand in small masses $\{ m_{\gamma} \}$ and small
external momenta $\{ q_{\gamma} \}$ of the subgraph $\gamma$ (before
integration); ``$ \circ $'' inserts the subgraph expansion in the
numerator of the integrand $F_{G/{\gamma}}$.  The sum goes over all
subgraphs $\gamma$ which (a) contain all lines with large masses, and
(b) are one-particle irreducible relative to lines with light or zero
masses.

The following types of integrals occur in the asymptotic expansion of
the muon AMM in the Standard Model:

\begin{enumerate}
\item
two-loop tadpole diagrams with various heavy masses on internal
lines;
\item
two-loop self-energy diagrams, involving contributions from fermions
lighter or heavier than the muon, with the external momentum on
the muon mass shell;
\item
two-loop self-energy diagrams with two or three muon  lines and
the external momentum on the muon  mass shell;
\item
various products of a one-loop self-energy diagram on shell and a one-loop
tadpole with a heavy mass.
\end{enumerate}

Almost all of these diagrams can be evaluated analytically using the
package SHELL2 \cite{SHELL2}. For our calculation we have modified
this package in the following way:

\begin{enumerate}
\item
There are no restrictions on the indices of the lines
(powers of scalar denominators).
\item
More recurrence relations are used, and the dependence on the space-time
dimension is always explicitly reducible to powers of linear factors.
\item
A new algorithm for simplification of this rational fractions is
implemented. These modifications
essentially reduce the execution time (in some cases, down to
the order of a hundredth).
\item
New programs for evaluating two-loop tadpole integrals with different
masses are added.
\item
New programs were written for the asymptotic expansion of one-loop
self-energy diagrams (relevant for renormalization)
in the large-mass limit.
\end{enumerate}

\section{The toy model}

As the first step, we concentrate on a ``toy'' model, a ``slice'' of
the Standard Model, involving a light charged spinor $\Psi$, the
photon $A_\mu$, and a heavy neutral scalar field $\Phi$.  The scalar
has triple $\left( g \right)$ and quartic $\left( \lambda \right)$
self-interactions, and the Yukawa coupling $\left( y \right)$ to the
spinor.  The Lagrangian of the toy model reads (in the Euclidean
space-time)

\begin{eqnarray}
{\it L} & = & \frac{1}{2} \partial_\mu \Phi \partial^\mu \Phi
+ \frac{1}{2} M^2 :\Phi^2: - \frac{g}{3!} :\Phi^3:
- \frac{\lambda}{4!} :\Phi^4:
+ \frac{1}{4} \left(\partial_\mu A_\nu - \partial_\nu A_\mu \right)^2
\nonumber \\ &&
+ \frac{1}{2 \alpha} \left( \partial_\mu A^\mu \right)^2
+ \bar{\Psi} \left( \hat{\partial} + m \right) \Psi
+ i e \bar{\Psi} \hat A \Psi - y \Phi \bar{\Psi} \Psi
\label{toy-model}
\end{eqnarray}

\noindent
where $e$ is the electric charge and $\alpha$ is a gauge fixing parameter.

The main aims of the present investigation are the following:

\begin{enumerate}
\item
Verification of the consistency of the large-mass asymptotic expansion
with the external momentum on the mass shell of a small mass.
In particular, we check the cancelation of all
threshold singularities that appear in individual diagrams
and manifest themselves as infrared poles in $\varepsilon$.
\item
Estimation of the influence of a heavy neutral scalar particle on the AMM
of the muon in the framework of the Standard Model.
\item
Verification of gauge independence (we use the covariant
gauge with an arbitrary parameter $\alpha$).
\end{enumerate}

\begin{figure}[t]
\vskip 80mm
\centerline{\vbox{\epsfysize=55mm \epsfbox{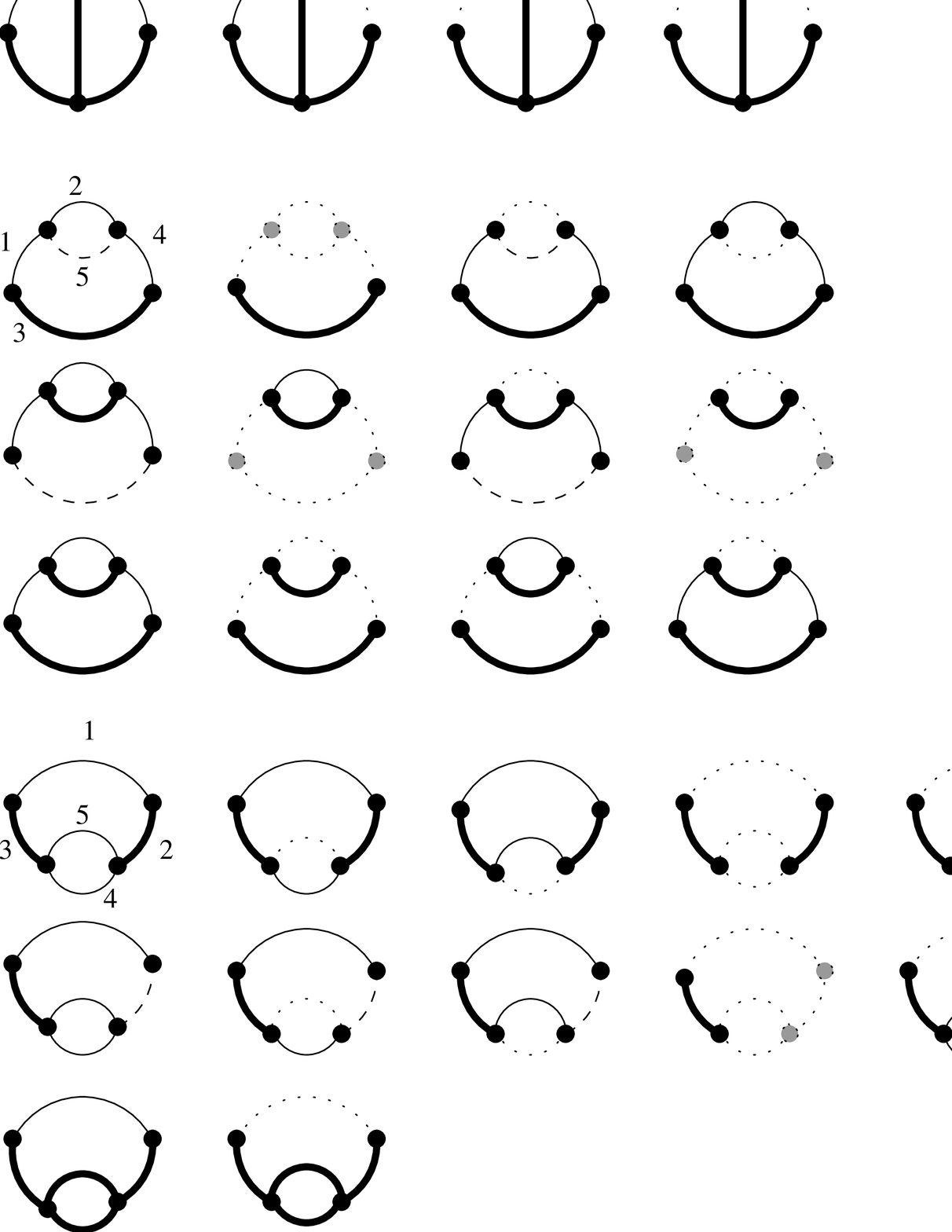}}}
\caption{\label{fig3} The prototypes and their subgraphs
contributing to the large mass expansion.
Thick, thin and dashed lines correspond to
the heavy-mass, light-mass, and massless propagators, respectively.
Dotted lines indicate the lines omitted in the subgraph $\gamma$.}
\end{figure}

In what follows we analyze in some detail the diagrams contributing
to the AMM of the fermion in our toy model and specify the
renormalization procedure.  Apart from counterterms $40$ diagrams
contribute to the two-loop AMM of the fermion.  After performing the
Dirac and Lorentz algebra, all diagrams can be reduced to some set of
scalar prototypes. A prototype defines the arrangement of massive and
massless lines in a diagram. Individual integrals are specified by
the powers of the scalar denominators, called indices of the lines.
From the point of view of the asymptotic expansion method the
topology of the diagram is essential. All diagrams of the toy model
that contribute to the two-loop AMM can be classified in terms of $9$
prototypes (we omit the pure QED diagrams).  These prototypes and
their corresponding subgraphs $\gamma$ involved in the asymptotic
expansion, are given in Fig.\ref{fig3}. In dimensional regularization, the
last subgraphs  vanish in cases $1, 4, 7$, and $8$, owing to massless
tadpoles.

\section{Program TLAMM}

All diagrams are generated in symbolic form by means of QGRAF
\cite{QGRAF}.  For automatic calculation  we have created a special
piloting program written in C.  This program, called TLAMM,

\begin{enumerate}
\item
reads QGRAF output;
\item
creates a file containing  the complete FORM program for calculating
each diagram;
\item
executes FORM;
\item
reads FORM output, picks out the result of the
calculation, and builds the total sum of all diagrams in a single
file which can be processed by FORM.
\end{enumerate}

\begin{figure}[ht]
\centerline{\vbox{\epsfysize=85mm \epsfbox{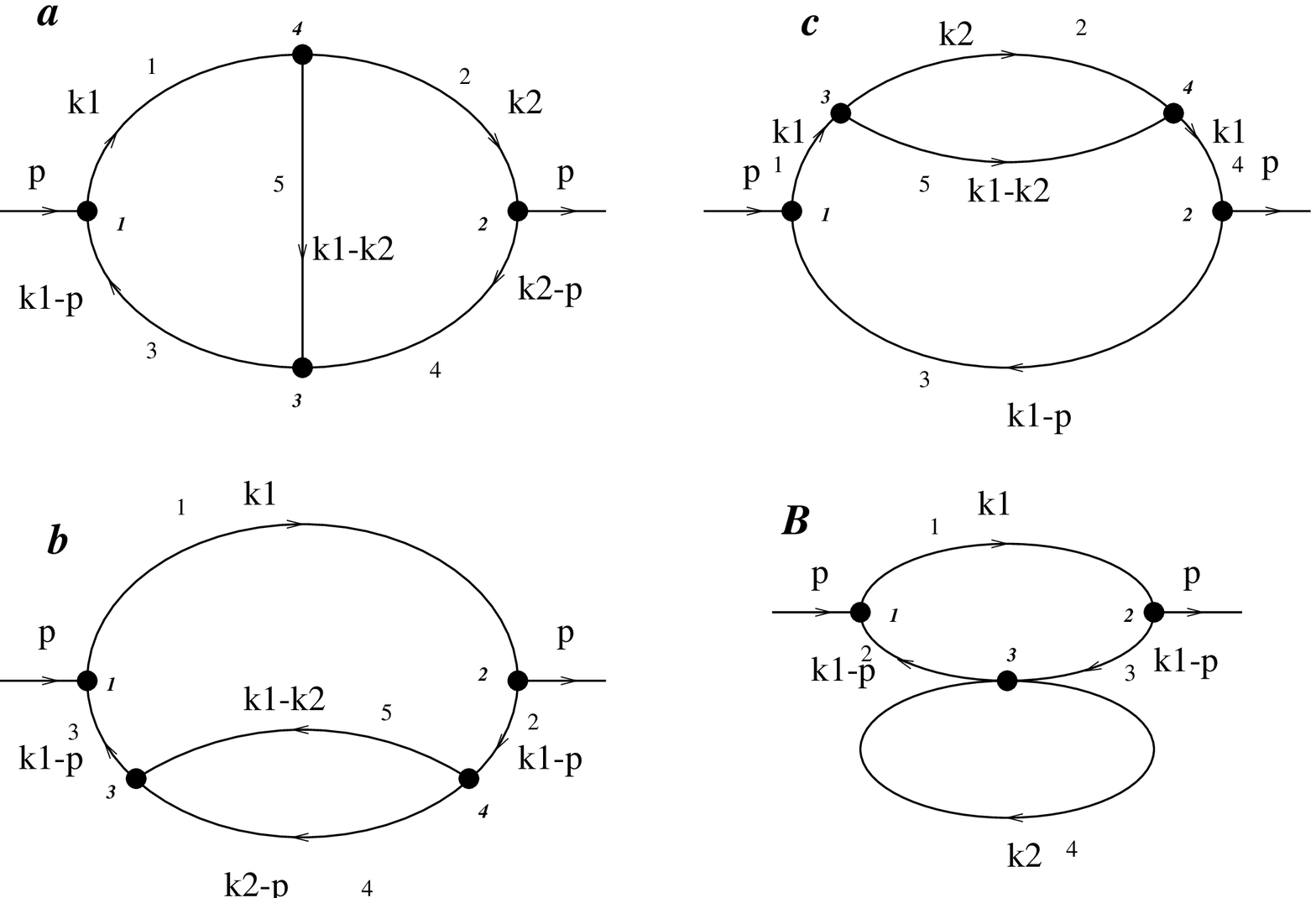}}}
\caption{\label{4top}Two-loop topologies existing for the AMM of the
lepton in the Standard Model.}
\end{figure}

The program has its own internal notation for topologies.  In the
problem $g-2$ of a lepton in the Standard Model there are four
different topologies (see Fig.\ref{4top}). Line number 1 is always assumed
to correspond to a fermion line (a lepton or a neutrino); therefore,
topologies {\tt b} and {\tt c} are distinguished.

\noindent
All diagrams are classified according to their {\tt local
prototypes}.  The notation consists of a letter (the topology) and
five (or four in {\tt B} case) integer numbers specifying the masses
on the lines

0 for the zero mass: $\gamma$, $\nu$;

1 for a mass less than $m_\mu$: $e, u, d$;

2 for $m_\mu$;

3 for an intermediate mass between $m_\mu$ and $M_W$: $s,\tau, c, b$;

4 for a heavy mass: $W, Z, t, H$.

\noindent
In the Feynman gauge $\alpha=1$ all pseudo-Higgs particles and the
Faddeev-Popov ghosts are heavy except for one massless ghost.

Each local prototype is calculated by means of the asymptotic
expansion in heavy masses. For each diagram the corresponding FORM
subroutine is called according to the local prototype.

Identifiers for vertices and propagators and the explicit Feynman
rules are red from separate files and then inserted into the FORM
program.  Because the number of identifiers needed for the
calculation of all diagrams together may exceed the FORM capacity,
the piloting program TLAMM retains for each diagram only those
involved in its calculation.

All initial settings are defined in a configuration file. The latter
contains information on the file names, identifiers of topologies,
the distribution of momenta, and the description of the model in
terms of the notation that is some extension of QGRAF's.
The program carries out the complete verification of all input files
except the QGRAF output.

There exist several options which allow one to process only the diagrams

\begin{itemize}
\item explicitly listed by number;
\item of a given prototype;
\item of a specified topology.
\end{itemize}

\noindent
There are also some debugging options.

The asymptotic expansion of each prototype is performed by a separate
FORM program. For efficiency of the algorithm the following points are
essential:

\begin{enumerate}
\item
The result of the calculation is presented as a series in small
parameters. Care is taken to avoid the production of unnecessarily high
powers in intermediate results.

\item
For the evaluation of the Feynman integrals, it is necessary to reduce scalar
products of momenta in the numerator to the square combinations which
are present in the denominator. Most efficiently this is done by means of
recurrence relations proposed by Tarasov \cite{Tarasov}.
\end{enumerate}

\begin{figure}[ht]
\vspace*{-52mm}
\centerline{\vbox{\epsfysize=285mm \epsfbox{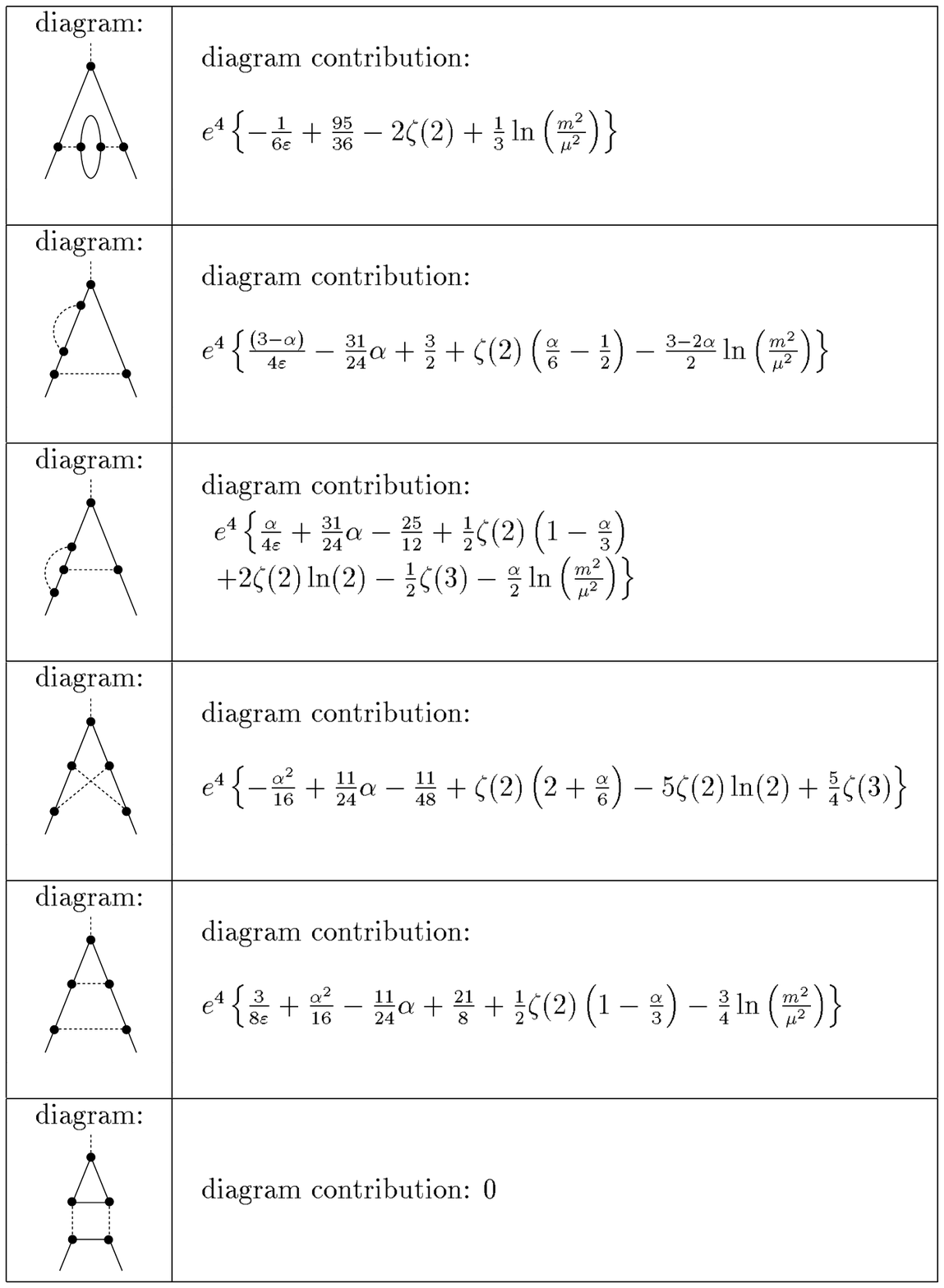}}}
\vspace*{-62mm}
\caption{\label{fig4}
The two-loop QED contributions to the AMM in the arbitrary gauge.}
\end{figure}

\begin{figure}[ht]
\vspace{-35mm}
\centerline{\vbox{\epsfysize=290mm \epsfbox{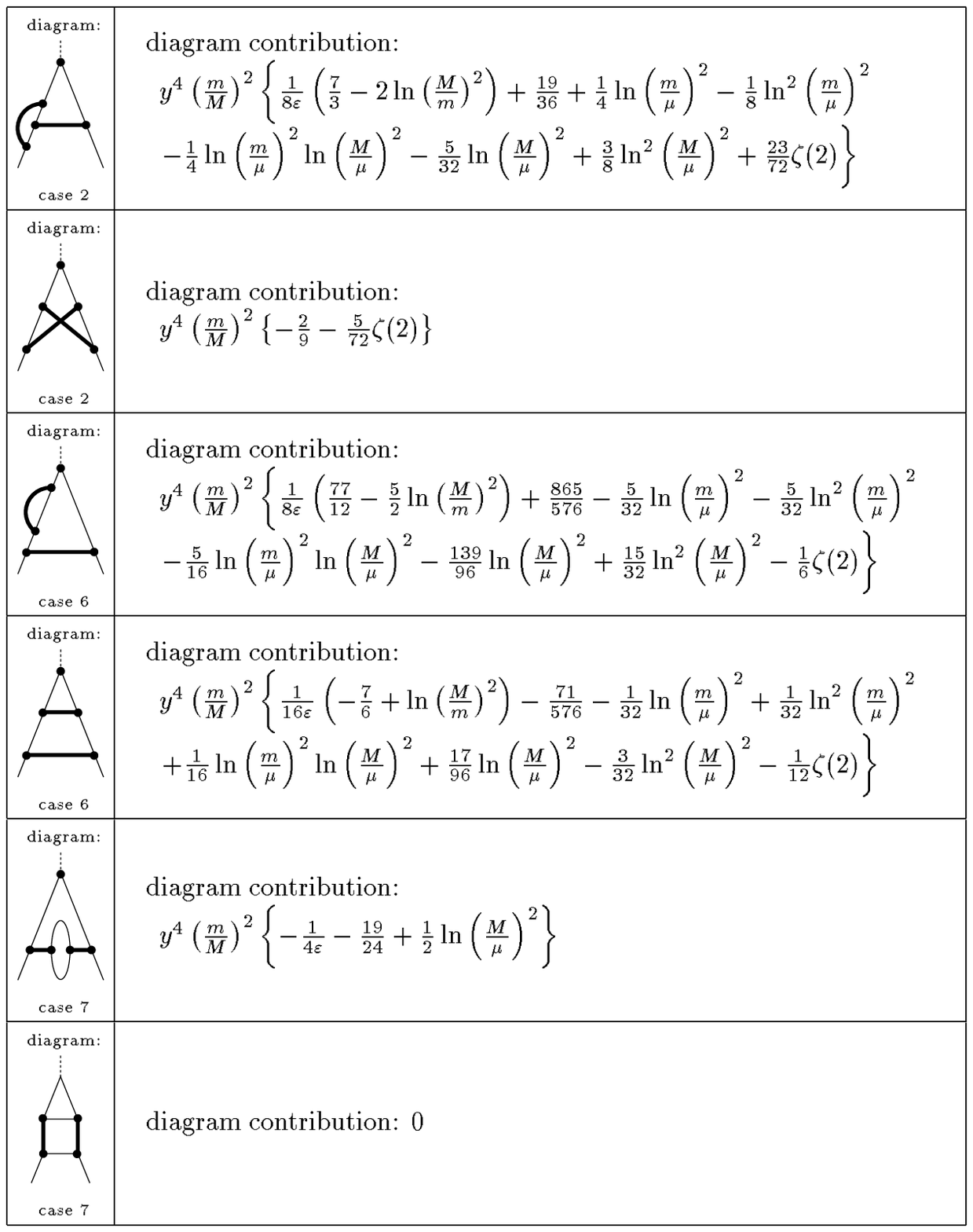}}}
\vspace{-80mm}
\caption{\label{fig5}
The two-loop AMM contributions proportional to $ y^4 $.}
\end{figure}

\begin{figure}[ht]
\vspace{-37mm}
\centerline{\vbox{\epsfysize=290mm \epsfbox{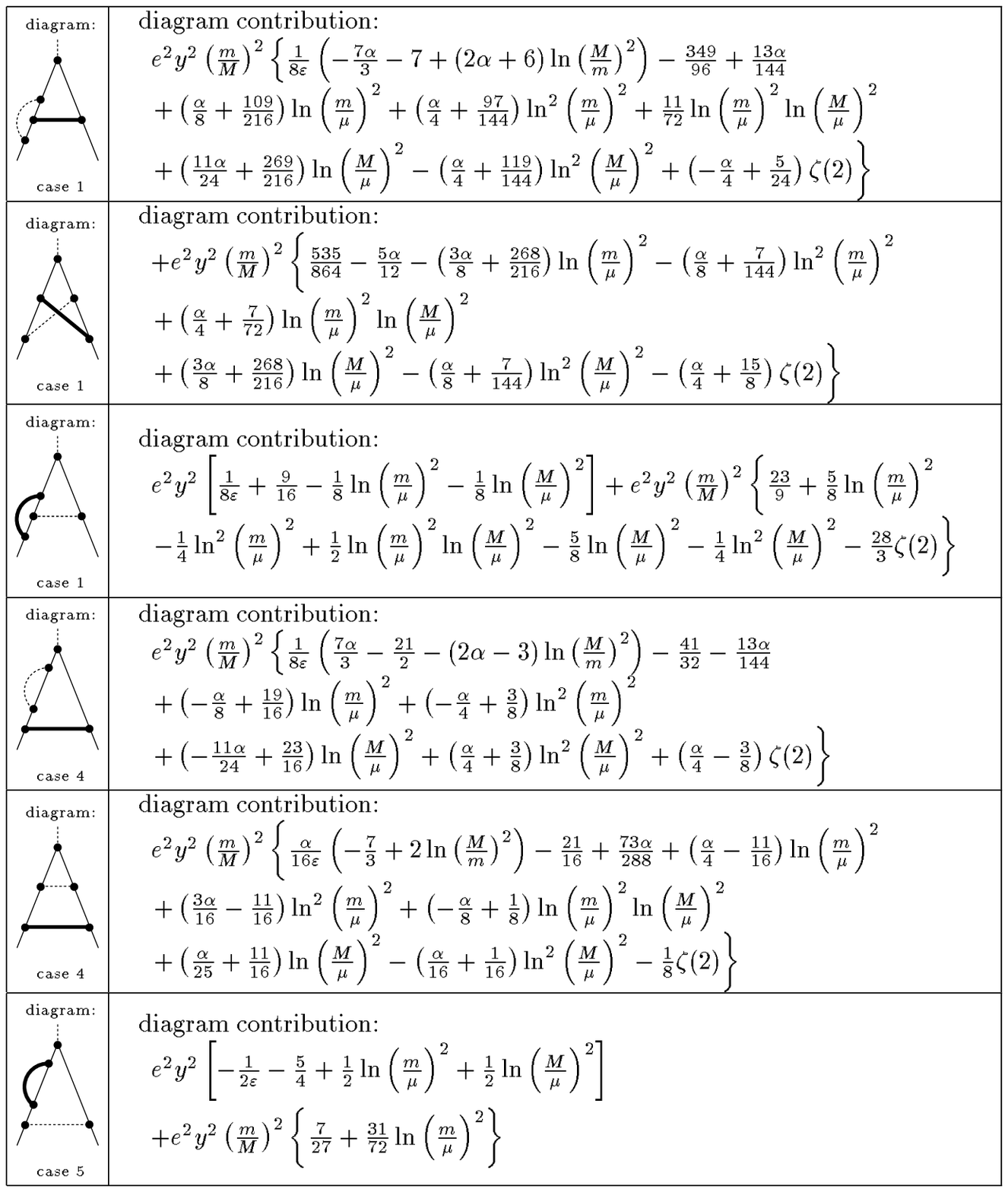}}}
\vspace{-82mm}
\caption{\label{fig6} The two-loop AMM contributions
proportional to $ e^2y^2 $ (to be continued).}
\end{figure}

\begin{figure}[htb]
\vbox{
\vspace*{-60mm}
\centerline{\vbox{\epsfysize=290mm \epsfbox{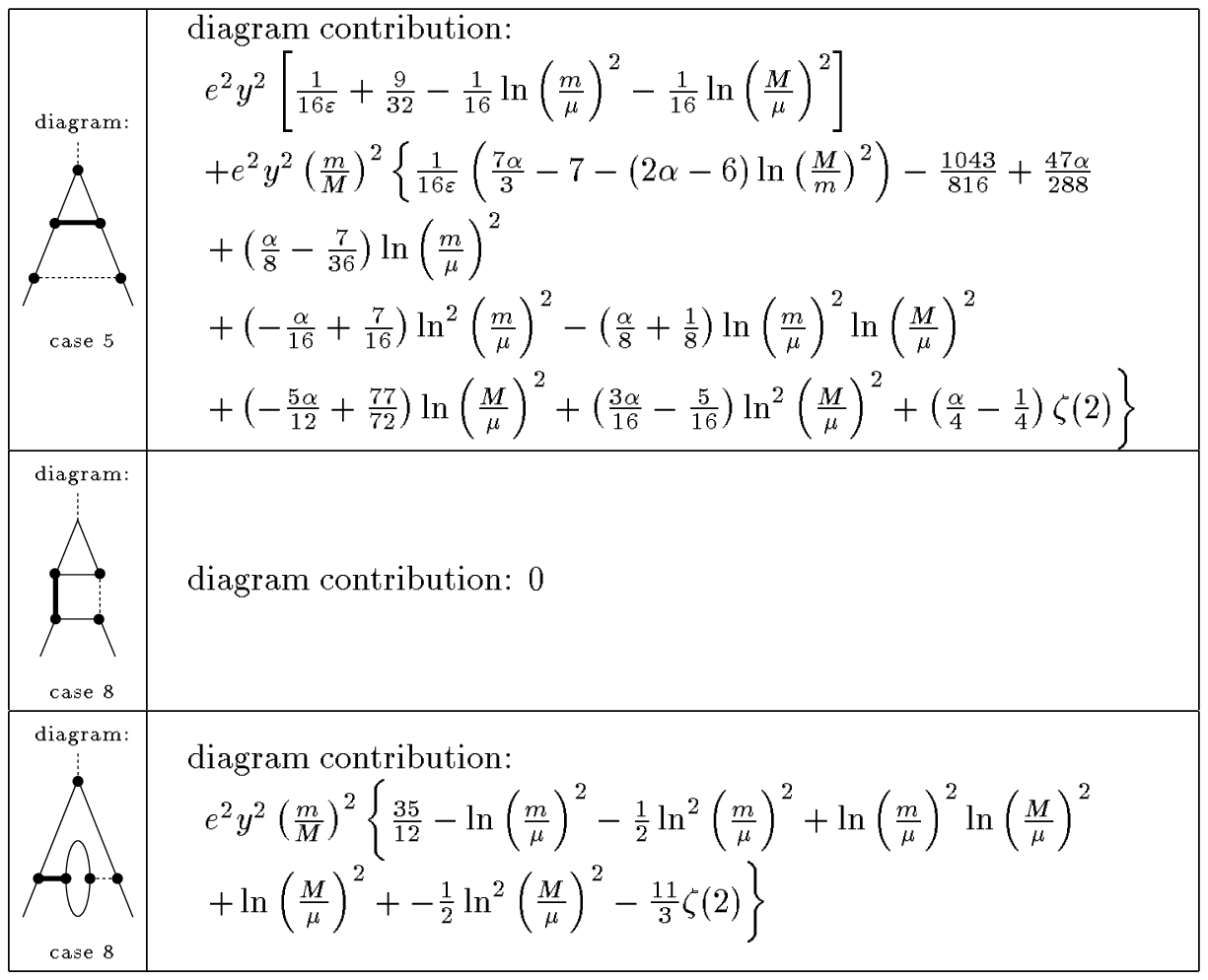}}}
\vspace*{-150mm}
}
\caption{\label{fig7}
The two-loop AMM contributions proportional to $ e^2y^2 $
(continued).}
\end{figure}

\begin{figure}[htb]
\vbox{
\vspace*{-45mm}
\centerline{\vbox{\epsfysize=290mm \epsfbox{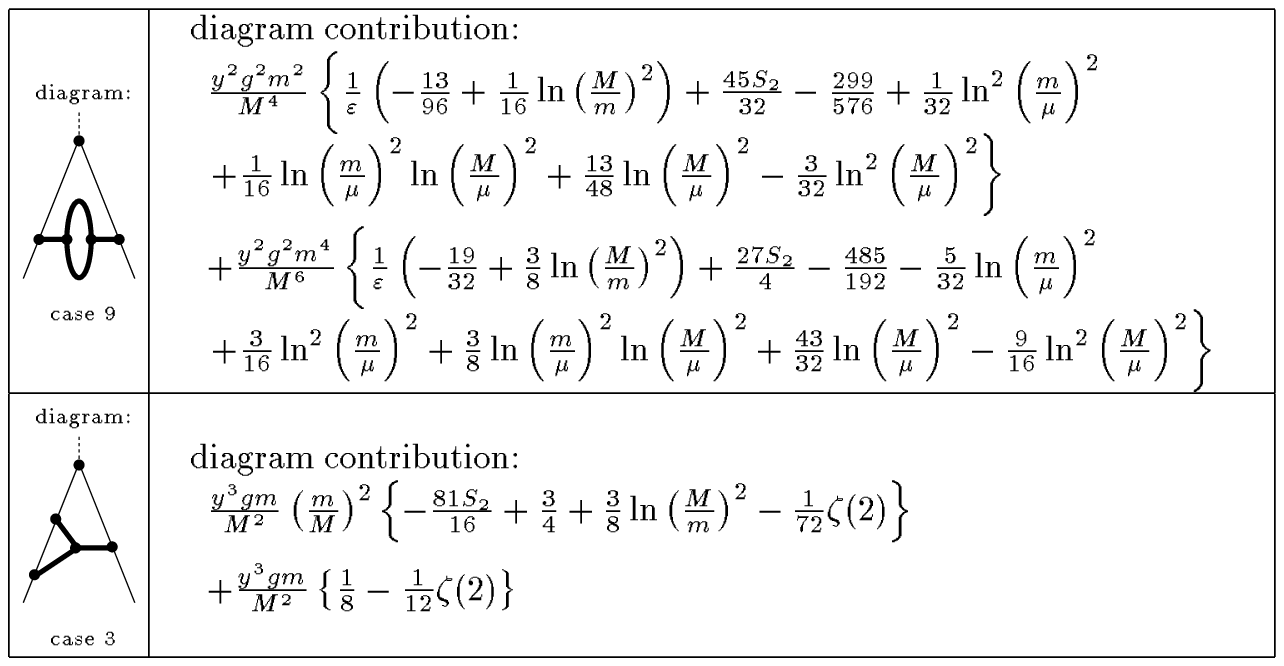}}}
\vspace*{-180mm}
}
\caption{\label{fig8}
The two-loop AMM contributions involving the triple interaction $ g $ .}
\end{figure}

As a demonstration of the functioning of TLAMM, the unrenormalized
(``bare'') contributions of all the two-loop diagrams to the anomalous
magnetic moment of the fermion in the toy model are presented in
Figs.\ref{fig4}--\ref{fig8} to the leading order in $m^2/M^2$. In
the presented expressions the factor of $(2 \pi)^{-4}$ is implied.
During the calculations, each loop was divided by
$\Gamma(1+\varepsilon)$ rather than multiplied by
$\Gamma(1-\varepsilon)$ as is done in the $\overline{\rm{MS}}$
definition.

\section{The results of the calculations}

Initially, we use the minimal subtraction scheme for the parameters of
the Lagrangian (\ref{toy-model}). Afterwards, it is more convenient to
re-express the running masses $m_R$ and $M_R^2$ in terms of the
physical pole masses $m, M^2$ of the particles
(at the one-loop level, relevant here),

\begin{eqnarray}
m_R (\mu^2 )
& = & m \Big \{ 1
- \frac{e_R^2}{16 \pi^2} \Big( 4 - 3 \ln \frac{m^2}{\mu^2} \Big)
+  \frac{y_R^2}{16 \pi^2} \Big[
\frac{5}{4} - \frac{3}{2} \ln \frac{M^2}{\mu^2}
\nonumber \\*
& + &
\left( \frac{m^2}{M^2} \right)
\Big( \frac{1}{6} - \ln \frac{M^2}{m^2}  \Big)
+ \left( \frac{m^2}{M^2} \right)^2
\Big( \frac{7}{8} - \frac{3}{2} \ln \frac{M^2}{m^2}
\Big)
\nonumber \\*
& + &
\left( \frac{m^2}{M^2} \right)^3
\Big( \frac{47}{20} - 3 \ln \frac{M^2}{m^2}  \Big)
+ \cdots \Big] \Big \},
\\
M_R^2 ( \mu ^2 ) & = & M^2 \Big \{ 1
+ \frac{y_R^2}{16 \pi^2} \Big[ 4 - 2 \ln \frac{M^2}{\mu^2}
- \left( \frac{m^2}{M^2} \right)
\Big( 16 - 12 \ln \frac{M^2}{\mu^2} \Big)
\nonumber \\*
& - &
\left( \frac{m^2}{M^2} \right)^2
\Big( 18 + 12 \ln \frac{M^2}{m^2} \Big)
 +
\left( \frac{m^2}{M^2} \right)^3
\Big( \frac{4}{3} - 8 \ln \frac{M^2}{m^2} \Big)
+ \cdots \Big]
\nonumber \\*
& + &
\frac{g_R^2}{16 \pi^2 M^2} \Big( 1 - \frac{\pi}{2 \sqrt{3}}
- \frac{1}{2} \ln \frac{M^2}{\mu^2} \Big)
\Big \},
\end{eqnarray}

\noindent
thus terming to the on-shell renormalization of the masses.
The scalar-field tadpoles do never contribute owing to the normal
ordering of the Lagrangian. Anyway, their contributions would become
of no consequence after everything is expressed in terms of the pole
masses. As a result, the quartic interaction constant $\lambda$ falls
out of the two-loop anomalous magnetic moment of the fermion.

We have calculated the asymptotic expansion up to the $7$th order in the
ratio $m^2/M^2$ and have convinced ourselves that all orders of the
expansion are free from on-shell singularities and involve only the
logarithms of the masses.  For brevity, we present just two leading
orders

\begin{eqnarray}
a_\mu & = &
\frac{e_R^2}{4 \pi^2} \left[ \frac{1}{2} \right]
+ \frac{e_R^4}{16 \pi^4} \left[
\frac{197}{144} +\left( \frac{1}{2}-3 \ln \left( 2 \right) \right)
\zeta \left(2 \right) + \frac{3}{4} \zeta \left(3 \right)
+ \frac{1}{6} \ln \frac{m^2}{\mu^2} \right]
\nonumber \\[3mm]
&&
+ \frac{y_R^2}{16 \pi^2}
\left(\frac{m}{M}\right)^2\left [
- \frac{7}{3}+2\ln \frac{M^2}{m^2}
\right ]
\nonumber \\[3mm]
&&
+\frac{e_R^2 y_R^2}{64 \pi^4}
\left(\frac{m}{M}\right)^2\left [
\frac{335}{27}
+ \frac{121}{9} \ln\frac{M^2}{\mu^2}
- \frac{179}{18} \ln\frac{m^2}{\mu^2}
\right.
\nonumber \\[3mm]
&&
\left.
-\frac{13}{2} \ln^2 \frac{M^2}{\mu^2}
-\frac{7}{2}\ln^2 \frac{m^2}{\mu^2}
+ 10 \ln \frac{M^2}{\mu^2} \ln \frac{m^2}{\mu^2}
-29 \zeta \left(2 \right)
\right ]
\nonumber \\[3mm]
&&
+\frac{y_R^3 }{256 \pi^4 } \left(\frac{g_R m}{M^2}\right)
\left [ 2  - \frac{4}{3} \zeta \left(2 \right)
 + \left(\frac{m}{M}\right)^2
\left\{\frac{46}{3}-\frac{189}{2}S_2
+6\ln \frac{M^2}{m^2}
\right\} \right ]
\nonumber \\[3mm]
&&
+\frac{y_R^2}{256 \pi^4}
\left(\frac{g_R m}{M^2}\right)^2\left [
- \frac{5}{3} + \frac{45}{2}S_2
-\frac{13}{6}\frac{\pi}{\sqrt{3}}
-\left(2-\frac{\pi}{\sqrt{3}}\right)
\ln \frac{M^2}{m^2}
\right ]
\nonumber \\[3mm]
&&
+\frac{y_R^4}{256 \pi^4}
\left(\frac{m}{M}\right)^2\left [
\frac{103}{6}
+13\ln \frac{m^2}{\mu^2}
-\frac{74}{3} \ln \frac{M^2}{\mu^2}
+10\ln \frac{M^2}{\mu^2} \ln \frac{M^2}{m^2}
\right ],~
\label{aR}
\end{eqnarray}

\noindent
where
$S_2 = \frac{4}{9 \sqrt{3}} {\rm Cl}_2
\left( \frac{\pi}{3} \right) = 0.2604341$
with ${\rm Cl}_2$ the  Clausen  function;
$e_R, y_R$ and $g_R$ are renormalized (running) coupling constants.

In quantum electrodynamics it is agreed to express the running charge
of the electron $e_R(\mu^2)$ in terms of the experimentally measurable
physical charge $e$. The later is defined by the nonrelativistic
Thompson limit of the Compton scattering, that is, as the product
of the on-shell vertex function at the zero momentum of the photon
[with the projection operator $\left(-i \hat{p} + m \right)$
 $\gamma_\mu$
$\left(-i \hat{p} + m \right)$~] by the on-shell wave-function
renormalization constant (the residue of the fermion propagator at
its pole). Both quantities are re\-norm\-al\-iz\-a\-tion-
and gauge-invariant
but contain an infrared singularity which cancels in the product.
Performing an analogous procedure for the charge of the muon in our
toy model to the one-loop order we get

\begin{equation}
e_R^2 (\mu^2 ) =
e^2 \left(1 - \frac{1}{3} \frac{e^2}{4 \pi^2} \ln \frac{m^2}{\mu^2}
\right).
\end{equation}

The physical Yukawa charge could also be defined in terms of the
on-shell Yukawa vertex. However, its evaluation is a difficult task in
itself. On the other hand, in the Standard Model the Yukawa charge is
usually related to the mass of the fermion rather than kept as an
independent parameter. As a compromise, to define the physical Yukawa
charge $y$, we use the following one-loop expression for the running
charge:

\begin{equation}
y_R^2 ( \mu^2 ) =
y^2 \left(1 + \frac{3}{2} \frac{e^2}{4 \pi^2} \ln \frac{m^2}{\mu^2}
- 5 \frac{y^2}{16 \pi^2} \ln \frac{M^2}{\mu^2} \right).
\end{equation}

\noindent
The calculation through the on-shell vertex would generally give a
finite correction (independent of $\mu^2$) to this formula.
The running of the triple scalar interaction is inessential in the
approximation that we consider.  In terms of the physical charges,
the anomalous magnetic moment (\ref{aR}) becomes

\begin{eqnarray}
a_\mu & = &
\frac{e^2}{4 \pi^2} \left[ \frac{1}{2} \right]
+ \frac{e^4}{16 \pi^4}\left[
\frac{197}{144} +\left( \frac{1}{2}-3 \ln  \left( 2 \right) \right)
\zeta \left(2 \right) + \frac{3}{4} \zeta \left(3 \right) \right]
\nonumber \\[3mm]
&&
+ \frac{y^2}{16 \pi^2}
\left(\frac{m}{M}\right)^2\left[
- \frac{7}{3}+2\ln \frac{M^2}{m^2}
\right]
\nonumber \\[3mm]
&&
+\frac{e^2y^2}{64 \pi^4}
\left(\frac{m}{M}\right)^2\left[
\frac{335}{27}+\frac{121}{9}\ln \frac{M^2}{m^2}
-\frac{13}{2}\ln^2 \frac{M^2}{m^2}
-29 \zeta \left(2 \right)
\right]
\nonumber \\[3mm]
&&
+\frac{y^3 }{256 \pi^4 } \left(\frac{g m}{M^2}\right)
\left[ 2  - \frac{4}{3} \zeta \left(2 \right)
 + \left(\frac{m}{M}\right)^2
\left\{\frac{46}{3}-\frac{189}{2}S_2
+6\ln \frac{M^2}{m^2}
\right\} \right]
\nonumber \\[3mm]
&&
+\frac{y^2}{256 \pi^4}
\left(\frac{g m}{M^2}\right)^2\left[
- \frac{5}{3}+\frac{45}{2}S_2
-\frac{13}{6}\frac{\pi}{\sqrt{3}}
-\left(2-\frac{\pi}{\sqrt{3}}\right)
\ln \frac{M^2}{m^2}
\right ]
\nonumber \\[3mm]
&&
+\frac{y^4}{256 \pi^4}
\left(\frac{m}{M}\right)^2\left[
\frac{103}{6}-13\ln \frac{M^2}{m^2}
\right].
\end{eqnarray}

The Standard-model motivated values for the Yukawa and triple scalar
interactions are

\begin{equation}
y = - \frac{1}{2} \frac{e}{\sin \Theta_W} \frac{m_\mu}{M_W},
\hspace*{15mm}
g = - \frac{e}{\sin \Theta_W} \frac{3 M_H^2}{2 M_W}.
\end{equation}

\noindent
Then, the estimate of the influence of a heavy neutral scalar particle
on the anomalous magnetic moment of the muon is
(besides the pure QED contribution)

\begin{eqnarray}
\Delta a_\mu & = &
\frac{e^2}{4 \pi^2} \frac{1}{\sin^2 \Theta_W}\left( \frac{m_\mu}{M_W}
\right)^2
\left(\frac{m_\mu}{M_H}\right)^2
\Biggl[
\Biggl(-\frac{7}{48}+\frac{1}{8}\ln\left(\frac{M_H}{m_\mu}\right)^2
\Biggr)
\nonumber \\[3mm]
& + &
\frac{e^2}{4 \pi^2}
\Biggl(
\frac{335}{432}+\frac{121}{144}\ln\left(\frac{M_H}{m_\mu}\right)^2
-\frac{13}{32}\ln^2\left(\frac{M_H}{m_\mu}\right)^2
-\frac{29}{16} \zeta \left(2 \right)
\Biggr)
\nonumber \\[3mm]
& + &
\frac{e^2}{4 \pi^2}
 \frac{1}{\sin^2 \Theta_W}\left( \frac{M_H}{M_W} \right)^2
\Biggl( -\frac{9}{256}  - \frac{1}{64} \zeta \left(2 \right)
+ \frac{405}{512}S_2
\nonumber \\*
& &
- \frac{39}{256} \frac{\pi}{\sqrt{3}}
+  \frac{9}{128} \left( \frac{\pi}{\sqrt{3}} - 1 \right)
\ln\left(\frac{M_H}{m_\mu}\right)^2
\Biggr)
\nonumber \\[3mm]
& + &
\frac{e^2}{4 \pi^2}
\frac{1}{\sin^2 \Theta_W}\left( \frac{m_\mu}{M_W} \right)^2
\Biggl(
\frac{379}{1536}-\frac{567}{512}S_2
+ \frac{5}{256} \ln\left(\frac{M_H}{m_\mu}\right)^2
\Biggr) \Biggr].
\end{eqnarray}

\noindent
This contribution is strongly suppressed and seems improbable to be
exhibited in future experiments.

We conclude that
\begin{enumerate}
\item
the gauge independence of the two-loop contribution to the anomalous
magnetic moment of the fermion has been verified;
\item
any threshold singularities in the on-shell asymptotic expansion
of the diagrams contributing to the AMM do cancel;
\item
the correction due to the heavy neutral scalar is
suppressed by the ratio of $m^4_\mu$ to heavy masses.
\end{enumerate}

\noindent
{\bf Acknowledgments}
This work was supported in part by the RFFI grant \# 96-02-17531,
by Volkswagenstiftung and by Bundesministerium
f\"ur Forschung und Technologie.

\end{document}